\documentclass{Interspeech}
\usepackage[T1]{fontenc}

\interspeechcameraready

\title{Regularizing Learnable Feature Extraction for Automatic Speech Recognition}

\author[affiliation={1}]{Peter}{Vieting}
\author[affiliation={1}]{Maximilian}{Kannen}
\author[affiliation={1,2}]{Benedikt}{Hilmes}
\author[affiliation={1,2}]{Ralf}{Schlüter}
\author[affiliation={1,2}]{Hermann}{Ney}

\affiliation{Machine Learning and Human Language Technology Group}{RWTH Aachen University}{Germany}
\affiliation{}{AppTek GmbH}{Germany}
\email{\{vieting,hilmes,schlueter,ney\}@ml.rwth-aachen.de}

\usepackage{comment}
\usepackage{siunitx}
\usepackage{multirow}
\usepackage{subcaption}
\usepackage{hyperref}
\usepackage{cleveref}
\usepackage[acronym,toc,shortcuts,nonumberlist]{glossaries}
\usepackage{xcolor}
\usepackage{xspace}
\usepackage{tikz}
\usepackage{pgfplots}
\usetikzlibrary{arrows,positioning}
\pgfplotsset{compat=1.18}

\newacronym{AM}{AM}{acoustic model}
\newacronym{AMI}{AMI}{Augmented Multi-party Interaction}
\newacronym{ARQ}{ARQ}{automatic repeat request}
\newacronym{ASR}{ASR}{automatic speech recognition}
\newacronym[longplural={bi-directional long-short term memories}]{BLSTM}{BLSTM}{bi-directional long-short term memory}
\newacronym{BSS}{BSS}{blind speech separation}
\newacronym{CART}{CART}{classification and regression tree}
\newacronym{CCA}{CCA}{canonical correlation analysis}
\newacronym{CE}{CE}{cross entropy}
\newacronym{CER}{CER}{character error rate}
\newacronym{CDp}{CDp}{context dependent phoneme}
\newacronym{CNN}{CNN}{convolutional neural network}
\newacronym{CPC}{CPC}{contrastive predictive coding}
\newacronym{CTC}{CTC}{connectionist temporal classification}
\newacronym{DCT}{DCT}{discrete cosine transform}
\newacronym{DL}{DL}{deep learning}
\newacronym{DNN}{DNN}{deep neural network}
\newacronym{DNN-HMM}{DNN-HMM}{deep neural network hidden Markov model}
\newacronym{ELBO}{ELBO}{evidence lower bound}
\newacronym{EM}{EM}{expectation maximization}
\newacronym{FE}{FE}{feature extractor}
\newacronym{FIR}{FIR}{finite impulse response}
\newacronym{FFNN}{FFNN}{feed-forward neural network}
\newacronym{fCE}{fCE}{frame-wise cross-entropy}
\newacronym{G2P}{G2P}{grapheme-to-phoneme conversion}
\newacronym{GAN}{GAN}{generative adversarial network}
\newacronym{GMM}{GMM}{Gaussian mixture model}
\newacronym{GMM-HMM}{GMM-HMM}{Gaussian mixture model hidden Markov model}
\newacronym{GPU}{GPU}{graphics processing unit}
\newacronym{HMM}{HMM}{hidden Markov model}
\newacronym{IHM}{IHM}{individual headset microphones}
\newacronym{IIR}{IIR}{infinite impulse response}
\newacronym{LAS}{LAS}{listen-attend-spell}
\newacronym{LC-BLSTM}{LC-BLSTM}{latency-controlled bidirectional long-short term memory}
\newacronym{LDA}{LDA}{linear discriminant analysis}
\newacronym{LM}{LM}{language model}
\newacronym[longplural={long-short term memories}]{LSTM}{LSTM}{long-short term memory}
\newacronym{MDM}{MDM}{multiple distant microphones}
\newacronym{MFCC}{MFCC}{Mel-frequency cepstral coefficients}
\newacronym{MHSA}{MHSA}{multi-head self-attention}
\newacronym{MRES}{MRES}{multi-resolutional}
\newacronym{MT}{MT}{machine translation}
\newacronym{NLP}{NLP}{natural language processing}
\newacronym{NN}{NN}{neural network}
\newacronym{OOV}{OOV}{out-of-vocabulary}
\newacronym{OCLR}{OCLR}{one-cycle learning rate}
\newacronym{PC2}{$\text{PC}^{\text{2}}$}{Paderborn Center for Parallel Computing}

\newacronym{RNA}{RNA}{recurrent neural aligner}
\newacronym{RNN}{RNN}{recurrent neural network}
\newacronym{RSAN}{RSAN}{recurrent selective attention network}
\newacronym{SAT}{SAT}{speaker adaptive training}
\newacronym{SDM}{SDM}{single distant microphone}
\newacronym{SDR}{SDR}{signal-to-distortion ratio}
\newacronym{SC}{SC}{supervised convolutional}
\newacronym{SCF}{SCF}{supervised convolutional features}
\newacronym{sMBR}{sMBR}{state-level minimum Bayes risk}
\newacronym{STFT}{STFT}{short time Fourier transform}
\newacronym{TDNN}{TDNN}{time delay neural network}
\newacronym[longplural={time-frequencies}]{tf}{tf}{time-frequency}
\newacronym{VAD}{VAD}{voice activity detection}
\newacronym{VGG}{VGG}{TODO needs to be added}
\newacronym{VTLN}{VTLN}{vocal tract length normalization}
\newacronym{cpWER}{cpWER}{concatenated minimum-permutation word error rate}
\newacronym{WER}{WER}{word error rate}
\newacronym{WERR}{WERR}{word error rate reduction}
\newacronym{WP}{WP}{work package}
\newacronym{WPE}{WPE}{weighted prediction error}
\newacronym{WSJ}{WSJ}{Wall Street Journal}
\newacronym{WSOLA}{WSOLA}{waveform similarity overlap-add}

\newcommand{\ms}[1]{\SI{#1}{\milli\second}}
\DeclareSIUnit\gigabyte{GB}
\begin{document}

\maketitle

\begin{abstract}
  Neural front-ends are an appealing alternative to traditional, fixed feature extraction pipelines for \gls{ASR} systems since they can be directly trained to fit the acoustic model.
  However, their performance often falls short compared to classical methods, which we show is largely due to their increased susceptibility to overfitting.
  This work therefore investigates regularization methods for training \gls{ASR} models with learnable feature extraction front-ends.
  First, we examine audio perturbation methods and show that larger relative improvements can be obtained for learnable features.
  Additionally, we identify two limitations in the standard use of SpecAugment for these front-ends and propose masking in the \gls{STFT}-domain as a simple but effective modification to address these challenges.
  Finally, integrating both regularization approaches effectively closes the performance gap between traditional and learnable features.
\end{abstract}
\glsresetall

\section{Introduction}
Feature extraction is a key component of any \gls{ASR} system.
Conventional methods rely on hand-crafted feature engineering which is often inspired by human auditory perception \cite{stevens1940mel,schlueter2007gammatone}.
While this has shown to be effective, these approaches risk discarding valuable speech information, potentially limiting the \gls{ASR} performance.
Learnable feature extraction methods address this limitation by optimizing features directly for the acoustic model and thus ensure a more tailored representation \cite{sainath2015learning,zeghidour2018learning,tueske2018waveform}.
Beyond the final performance, the use of learnable features aligns with the broader goal of designing end-to-end ASR systems, where the entire processing pipeline -- from raw waveform to predicted labels -- is optimized jointly within a single monolithic neural framework.

\begin{figure}[htb]
    \centering
    \input{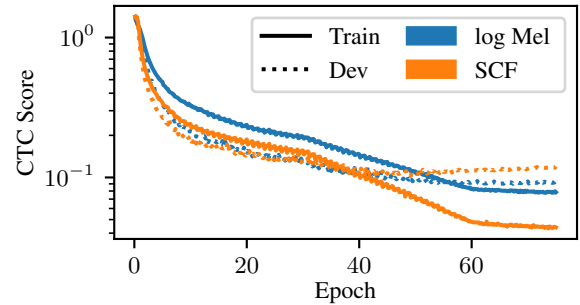}
    \vspace{-0.4em}
    \caption{Train and dev connectionist temporal classification (CTC) scores for the baseline training with log Mel and supervised convolutional features (SCF), respectively, demonstrating the overfitting issue for the latter.}
    \label{fig:scores}
\end{figure}

In practice, learnable feature extraction front-ends often struggle to compete with classical methods especially when the amount of training data is limited \cite{tueske2018waveform}.
While prior works have not explicitly considered overfitting as a problem, we hypothesize that learnable front-ends are more susceptible to it.
\Cref{fig:scores} clearly shows this effect.
This makes overfitting an important concern that causes a performance gap.
There are well-known methods to mitigate it during neural network training in general \cite{hanson1988weightdecay, srivastava2014dropout}.
However, their application to learnable feature extractors in \gls{ASR} has received little attention.
In this work, we investigate perturbations of the input audio during training as an effective regularization strategy to improve generalization.
Furthermore, we demonstrate that the default use of SpecAugment \cite{park2019specaugment} is suboptimal for learnable feature extractors and propose a refined method to address these limitations.

Naturally, overfitting is particularly challenging when training data is limited.
In this study, we focus on scenarios with substantially less data than typically used in previous studies that achieve competitive performance with learnable front-ends \cite{sainath2015learning}.
The Switchboard corpus with \SI{311}{\hour} of data is a suitable choice for this.
Beyond the size, this corpus is particularly challenging for learnable features because of its telephony speech nature with a limited bandwidth of \SI{8}{\kilo\hertz}.
Our approach is thus viable for demanding low resource scenarios and situations where large pretrained models are prohibitive due to efficiency constraints.

The key contributions of this work are as follows:
\begin{itemize}
  \item We show that overfitting is a problem when training \gls{ASR} models with learnable features.
  \item A systematic study on the effect of audio perturbation for learnable front-ends is presented which, to the best of our knowledge, has not been explored before.
  \item A novel variant of SpecAugment is proposed to specifically address the challenges of learnable feature extraction.
  \item We demonstrate that combining both regularization methods allows the learnable front-end to perform on par with traditional features on Switchboard with \SI{311}{\hour} of training data.
\end{itemize}

\vspace{-3mm}

\section{Related Work}
While there exists a line of research on learnable front-ends for \gls{ASR} and related tasks \cite{sainath2015learning,zeghidour2018learning,tueske2018waveform,zeghidour2021leaf}, the mitigation of overfitting in this context has not been addressed so far.
Parametric methods \cite{ravanelli2018sincnet} may inherently suffer less from overfitting due to their restricted modeling capacity.
However, this limitation may potentially affect the overall performance.
As an alternative approach, we therefore aim to use regularization methods during the neural network training to mitigate overfitting in this work.

The literature provides a rich set of techniques for regularizing neural network training such as weight decay \cite{hanson1988weightdecay}, dropout \cite{srivastava2014dropout} or label smoothing \cite{szegedy2016llabelsmoothing}.
Perturbation of the input is another important regularization technique.
Common methods for audio perturbation are speed, tempo and pitch perturbation \cite{ko2015augmentation,jaitly2013vtlp,kanda2013distortion}.
Further ideas to augment the audio input include signal companding \cite{das2021mulaw} and mixup \cite{medennikov2018mixup}.

Additionally, SpecAugment \cite{park2019specaugment} has been shown to be particularly effective for \gls{ASR}.
It works by masking blocks of time frames or feature channels during training as well as performing time warping.
Following its success, numerous variations have been proposed, e.g. dropping \cite{jain2022spliceout} or swapping \cite{song2020specswap} blocks instead of masking, masking based on phonemes or other semantic regions \cite{wang2020semanticmask,zhang2022phoneme_masking} and more \cite{wang2021specaugpp,damania2022augmentation,han2023randmasking}.
\cite{luo2024specaug_w2v} proposes a variation of SpecAugment's time warping for a wav2vec 2.0 model with learnable front-end.
However, this modification does not appear to be specifically designed for learnable feature extraction and the experimental results do not isolate its individual contribution to the performance improvement.

The SpecAugment variation proposed in this work is different from previous work in that it works in the \gls{STFT}-domain and caters specifically for the needs of learnable front-ends.

\section{Methods}
\subsection{Feature Extraction}
\label{sec:feature_extraction}
Typically, acoustic models for \gls{ASR} operate on features that represent the input speech.
One of the most common representations are log Mel filterbank features.
They are obtained by first computing the \gls{STFT}, in our case with window size \ms{25} and shift \ms{10}, and then taking the square of the magnitude.
Using Mel warping, an 80-dim vector is obtained per frame and finally a logarithmic compression and normalization are applied.

As the learnable feature extraction front-end, we use supervised convolutional features similar to \cite{tueske2018waveform,vieting2021waveform,vieting2023itg}.
The main components are two convolutional layers that operate as time-frequency-decomposition and temporal integration.
These operations resemble the functional principle of the Gammatone filterbank and the Hanning window used for Gammatone features \cite{schlueter2007gammatone}.
However, all filters are randomly initialized and trained with the rest of the neural network with the same supervised objective function.
The first convolutional layer takes the waveform as input.
It has 150 filters with a size of \ms{16} and a stride of \ms{0.625}.
The second layer has 5 filters with a size of 40 frames and a stride of 16, resulting in feature frames with a \ms{10} shift.
Since each of the 5 filters is applied to all 150 output channels of the first layer, the final feature dimension is 750.
In contrast to Gammatones features, a multi-resolutional temporal integration is achieved because 5 different filters can be learned.
Additionally, we use the absolute value as the element-wise activation function after the first layer and the 2.5\textsuperscript{th} root of the absolute value after the second layer.
Finally, layer normalization is applied.
In this case of a learnable front-end, the border between feature extraction and acoustic model blurs and the neural network operates directly on the raw waveform.

Both feature extraction blocks are followed by SpecAugment (see \Cref{sec:SpecAugment}) and VGG convolutions, resulting in a subsampling by a factor of 4 and a frame shift of \ms{40}.

\subsection{Audio Perturbation}
\label{sec:perturbation}
For our work, we investigate multiple ways of modifying the audio signal in order to regularize the training of our model.
Each of the perturbations is applied at the sequence level with a specified probability $p$, where $p < 1$ ensures to also include original audio in training.
While most earlier works generate copies of the corpus with fixed perturbation factors, we sample random factors for the modification in a given range on-the-fly to increase the variability of perturbed examples.
This allows the same sequence to pass through different modifications across multiple training iterations.

\textbf{Speed, Tempo and Pitch Perturbation}: The most common way of modifying an audio signal is changing its speed, tempo or pitch.
For speed perturbation, we resample the audio by a factor $a$.
This results in both the duration and the pitch being modified.
Tempo perturbation only changes the tempo and thus the duration of the audio while keeping the pitch level constant.
This is achieved using the \gls{WSOLA} algorithm \cite{verhelst1993wsola}.
Lastly, we can also only modify the pitch of the audio while keeping the duration constant.
Again we use \gls{WSOLA}, by first changing the tempo of the audio signal but then resample the signal back to the original duration.

\textbf{Nonlinear Amplitude Perturbation}:
The next perturbation involves a nonlinear distortion of the audio signal's amplitudes.
It is defined as:
\begin{equation}
  \tilde{x}(t) = \text{sign}(x(t)) \cdot | x(t) | ^ \beta
\end{equation}
where $ x(t)$ is the original audio normalized to be between -1 and 1, $\beta$ is the factor that controls the strength of the distortion and $\tilde{x}(t)$ is the perturbed audio.
The perturbation either results in a more peaky signal or de-emphasizes outliers.

\textbf{\boldsymbol{$\mu$}-Law Perturbation}:
The next perturbation technique we investigate is based on $\mu$-law companding.
This companding algorithm reduces the dynamic range of a given signal and is typically used in telecommunication systems.
However, previous work has also applied it as a method for speech data augmentation \cite{das2021mulaw}.
We use the continuous encoding equation which offers the possibility to control the strength of the augmentation effect by a single parameter $\mu$:
\begin{equation}
  \tilde{x}(t) = \text{sgn}(x(t)) \cdot \frac{\ln(1 + \mu \cdot | x(t)|)}{\ln(1 + \mu)}
\end{equation}
While in telecommunication usually $\mu$ is set to 255, we use significantly smaller values in order to avoid perturbed examples that deviate too far from the distribution of the real training data.
Note that since $\mu > 0$, this is the only perturbation where in general $E[\tilde{x}(t)] \neq x(t)$.

\textbf{Preemphasis}:
Preemphasis is a filter technique often used for \gls{ASR} to emphasize higher frequencies in the audio signal.
The filter output $y(t)$ is defined as:
\begin{equation}
  \label{eq:preemphasis}
  y(t) = x(t) - \alpha x(t - 1)
\end{equation}
where $\alpha$ is usually chosen to be close to 1.
To avoid any mismatch, preemphasis is applied in the same way during training and inference.
In addition to the standard usage of preemphasis, we experiment with a second level of preemphasis as a perturbation technique.
For this, we apply Eq.~(\ref{eq:preemphasis}) a second time with a random value $\tilde{\alpha}$ sampled from a pre-defined range close to 0.
For inference, we only apply the regular preemphasis with $\alpha$.

\subsection{SpecAugment}
\label{sec:SpecAugment}
SpecAugment \cite{park2019specaugment} is a regularization technique that involves masking random regions of the extracted features along both the time and feature dimensions during training.
Similar to dropout \cite{srivastava2014dropout}, this approach prevents overfitting by ensuring the model does not overly depend on any particular part of the features.
However, unlike dropout, SpecAugment introduces structure by masking contiguous regions of data.
Specifically, when applying time masking, blocks of consecutive time frames are masked.
Since neighboring frames typically have a high correlation, masking a single frame results in minimal information loss.
In contrast, masking entire regions pushes the model to learn dependencies over longer contexts.

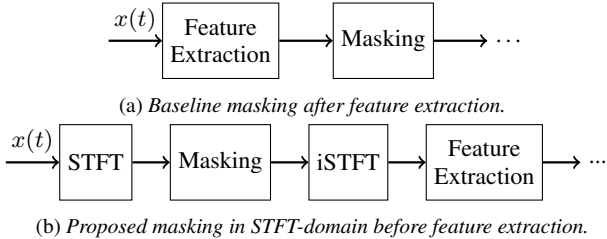
\begin{figure}
    \centering
    \begin{subfigure}[b]{\linewidth}
        \centering
        \begin{tikzpicture}[auto]
            \node[rectangle, minimum height=1.00cm, align=center] (xt1) at (0, 0) {};
            \node[rectangle, minimum height=1.00cm, align=center, draw, right=20 pt of xt1] (fe1) {Feature \\ Extraction};
            \node[rectangle, minimum height=1.00cm, align=center, draw, right=20 pt of fe1] (masking1) {Masking};
            \node[rectangle, minimum height=1.00cm, align=center, right=20 pt of masking1] (am1) {\dots};
            \draw [->, line width=0.3mm] (xt1) -- (fe1) node[midway, above, black] {$x(t)$};
            \draw [->, line width=0.3mm] (fe1) -- (masking1);
            \draw [->, line width=0.3mm] (masking1) -- (am1);
        \end{tikzpicture}
        \caption{Baseline masking after feature extraction.}
        \label{fig:masking_baseline}
    \end{subfigure}
    \begin{subfigure}[b]{\linewidth}
        \centering
        \hspace*{-10pt}
        \begin{tikzpicture}[auto]
            \node[rectangle, minimum height=1.00cm, align=center] (xt2) at (0, 0) {};
            \node[rectangle, minimum height=1.00cm, align=center, draw, right=20 pt of xt2] (stft) {STFT};
            \node[rectangle, minimum height=1.00cm, align=center, draw, right=14 pt of stft] (masking2) {Masking};
            \node[rectangle, minimum height=1.00cm, align=center, draw, right=14 pt of masking2] (istft) {iSTFT};
            \node[rectangle, minimum height=1.00cm, align=center, draw, right=14 pt of istft] (fe2) {Feature \\ Extraction};
            \node[rectangle, minimum height=1.00cm, align=center, right=14 pt of fe2] (am2) {...};
            \draw [->, line width=0.3mm] (xt2) -- (stft) node[midway, above, black] {$x(t)$};
            \draw [->, line width=0.3mm] (stft) -- (masking2);
            \draw [->, line width=0.3mm] (masking2) -- (istft);
            \draw [->, line width=0.3mm] (istft) -- (fe2);
            \draw [->, line width=0.3mm] (fe2) -- (am2);
        \end{tikzpicture}
        \caption{Proposed masking in STFT-domain before feature extraction.}
        \label{fig:masking_stft}
    \end{subfigure}
    \caption{Baseline vs. proposed masking strategies.}
    \label{fig:masking}
\end{figure}

For log Mel features, this logic trivially transfers to masking in the feature dimension since neighboring feature channels correspond to adjacent frequency regions.
However, this is fundamentally different in the case of learnable features.
While the model also typically learns a set of bandpass filters \cite{tueske2018waveform,vieting2023itg}, it cannot be assumed that adjacent channels represent similar frequencies since the order of filters is arbitrary.
We hypothesize that this limits the efficacy of SpecAugment for learnable front-ends.
Furthermore, SpecAugment is applied after the feature extraction as depicted in \Cref{fig:masking_baseline}.
This causes the gradient for the masked positions to be 0, resulting in fewer updates for each front-end filter.
As another consequence, learnable layers before and after the masking allow the model to shift information and bypass the masks resulting in ineffective regularization.

In this work, we investigate two different ways to solve these issues.
The first way is to sort the filters based on the peak frequency in their frequency response and to mask channels with neighboring peak frequencies.
However, this is complicated by the fact that the learned filters are not perfectly sharp bandpass filters but sometimes include wide or even multiple passbands \cite{tueske2018waveform, vieting2023itg}.
Thus, a proper masking of frequency areas cannot be guaranteed.
Also, the masking is still applied after the feature extractor such that this approach only addresses the first issue described above.

For the second proposed approach, we apply SpecAugment before the feature extraction process.
We apply masking in both time and frequency dimension in the \gls{STFT}-domain, transform the signal back using the inverse \gls{STFT} and use the resulting signal as input to our model.
This way, we can apply SpecAugment before feature extraction and guarantee that both time and frequency areas are masked properly, tackling both issues described above.
The information flow for the baseline and proposed masking strategies is visualized in \Cref{fig:masking}.

\section{Experimental Setup}
\subsection{Data}
The Switchboard-1 Release 2 dataset \cite{godfrey1992switchboard} serves as training data for our experiments.
The corpus contains \SI{311}{\hour} of conversational English telephony speech sampled at \SI{8}{\kilo\hertz}.
We use Hub5'00 as the development set and Hub5'01 for evaluation.

\subsection{ASR Model}
For our experiments, we use a \gls{CTC}-based \gls{ASR} model \cite{graves2006ctc}.
We follow the setup from a previous work \cite{zhou2022efficient} and use a 4-gram \gls{LM} during recognition.
Feature extraction is done as explained in \Cref{sec:feature_extraction} including VGG-style subsampling, followed by dropout and a final linear layer producing representations with a shift of \ms{40}.
Following results from preliminary experiments, standard preemphasis with $\alpha = 0.97$ is used for the supervised convolutional features, but not for log Mel.
Also, for the log Mel baseline, it was beneficial to normalize the input waveform over time and the resulting features over the batch.
Our encoder consists of 12 conformer \cite{gulati2020conformer} layers with a hidden dimension of 512.
Compared to the original paper, we swap the position of convolution and multi-head self-attention modules.
A final linear layer maps the encoder output to the target vocabulary, resulting in 74M parameters for the log Mel model.
When using learnable features, there are around 21k parameters for the feature extractor and the number of parameters for the linear layer after VGG increases from 1.3M to 12.3M because of the almost 10 times larger feature dimension\footnote{The VGG output dimension is 32 times the feature dimension resulting in $750\cdot32\cdot512\approx12.3\text{M}$ parameters for the linear layer.} leading to 85M parameters in total.
We train our model using the NAdam optimizer applying L2 weight decay \cite{hanson1988weightdecay} with a factor of 0.01 for the VGG frontend and 0.0001 for the encoder.
Further, dropout \cite{srivastava2014dropout} is applied with a probability of 0.1 for regularization.
We split the data into 6 sub-epochs and train a total of 450 sub-epochs, evaluating our model on sub-epochs 400 to 450 in steps of 10 and select the best checkpoint based on \gls{WER} on Hub5'00 for test evaluation.
We apply SpecAugment from the first sub-epoch.
To increase training stability, perturbations are only enabled after the first 4 full epochs by initializing the model with the corresponding checkpoint of the baseline.
To keep the total number of training epochs constant, we stop training with perturbations 4 full epochs earlier.
We use a \gls{OCLR} schedule starting from $1.325 \cdot 10^{-5}$ with a peak of $4.0 \cdot 10^{-4}$ after 180 sub-epochs going down to $1.0 \cdot 10^{-5}$ over another 180 sub-epochs and staying at that level for the final sub-epochs of training.
The experiments can be run on a single consumer \gls{GPU} with \SI{24}{\gigabyte} VRAM (e.g. Nvidia RTX 3090) so that the barrier for reproduction is low.
Also, the software is publicly accessible \cite{zeyer2018returnn,zhou2023rasr2} and the recipes are available online\footnote{\url{https://github.com/rwth-i6/returnn-experiments/tree/master/2025-regularizing-learnable-features}}.

\section{Experimental Results}
\subsection{Audio Perturbation}
We study the effect of the audio perturbation techniques introduced in \Cref{sec:perturbation} in \Cref{table:results_perturbations}.
Due to hardware availability constraints, we run the ablation study for \Cref{table:results_perturbations} on less powerful \glspl{GPU} with only \SI{11}{\gigabyte} of memory.
To facilitate this, the batch size is reduced from \SI{100}{\second} to \SI{50}{\second} and the gradients are accumulated over two batches.
Note that this is not strictly equivalent when the model contains batch normalization.
Additionally, preemphasis with $\alpha = 1.0$ yielded better results for some trainings and we always report the lower \gls{WER} in \Cref{table:results_perturbations}.
The full ablation study contains a large number of experiments, however, in the interest of clarity, we present only those experiments here that deliver the best results for log Mel and supervised convolutional features for each perturbation type.

For each experiment, we define the probability $p$ of perturbing the audio input as well as the range to sample the hyperparameter controlling the perturbation strength from.
We check $p \in \{0.3, 0.7, 1.0\}$ for most perturbations to cover a wide range of probabilities.
For speed perturbation, we sample from the range that is typically used, i.e.,  $a \sim U(0.9,1.1)$ \cite{ko2015augmentation}, but also from the less common $a \sim U(0.88,1.12)$ \cite{chen2020dataaugmentationchildrensspeech}.
Most other works use a range of $[0.9, 1.1]$ for tempo perturbations as well \cite{ko2015augmentation}.
However, our experiments showed that larger ranges are more beneficial and therefore our best results with tempo perturbation are obtained with $a \sim U(0.7,1.3)$.
For pitch perturbation, a change of -2 to 2 semitones roughly corresponds to speed changes of 0.9 to 1.1 and is therefore our starting point.
Yet, larger ranges degraded the performance.

For non-standard perturbations, selecting appropriate hyperparameters is more challenging due to the lack of prior references.
To address this, we determined the minimum and maximum perturbation factors such that the audio still sounded natural based on the authors' subjective perception.
Based on initial experimental results, we further tuned the parameters.

\Cref{table:results_perturbations} shows the obtained results.
It is clear that tempo perturbation achieves the best \glspl{WER} across both feature types.
While the log Mel model also clearly benefits from the nonlinear amplitude, $\mu$-law and preemphasis perturbation, only the latter yields a small improvement for the learnable features.
However, the overall best results exhibit a larger relative improvement for the learnable front-end.
\vspace{-2mm}

\begin{table}[htbp]

\centering
\caption{Ablation study of different audio perturbation methods. Results are reported for both log Mel and supervised convolutional features (SCF) on Hub5'00. All results use standard SpecAugment.}
\label{table:results_perturbations}
\vspace{-3mm}
\begin{tabular}{|c|c|c|c|S[table-format=2.1]|S[table-format=2.1]|}
\hline
                \multicolumn{4}{|c|}{Perturbation} & \multicolumn{2}{c|}{{WER [\%]}} \\\hline
        Type &                  $p$ &   Min &  Max & {log Mel} &               {SCF} \\\hline\hline
        None &                    - &     - &    - &      13.0 &                13.2 \\\hline
       Speed &                  0.7 &  0.88 & 1.12 &      13.1 &                13.3 \\\hline
       Tempo &                  1.0 &   0.7 &  1.3 &      12.6 &                12.6 \\\hline
       Pitch & \multirow{2}{*}{0.7} &    -2 &    2 &      13.2 &                13.4 \\\cline{1-1}\cline{3-6}
Nonlin. Amp. &                      &   0.8 &  1.2 &      12.6 &                13.2 \\\hline
   $\mu$-law &                  0.3 &     1 &    5 &      12.6 &                13.2 \\\hline
 Preemphasis &                  0.7 & -0.05 & 0.05 &      12.7 &                13.0 \\
\hline
\end{tabular}

\end{table}

\vspace{-5mm}

\subsection{SpecAugment Variations}
\label{sec:results_specaug}
The next set of experiments deals with the analysis of the previously proposed variations of SpecAugment.
A maximum time mask size of 15 frames was tuned for the log Mel baseline model and previously copied for experiments with the learnable front-end.
The maximum feature mask sizes were tuned individually for each row in \Cref{table:results_specaug_variants_bs10k_v2} and are 8 or 15 in the baseline cases.
Now, we first evaluate creating larger time masks while reducing the number of masks at the same time to keep the ratio of masked time frames constant.
Because of the temporal correlation of the input, this has a stronger regularization effect and forces the model to learn longer context dependencies.
As visible in \Cref{table:results_specaug_variants_bs10k_v2}, this has little effect for the log Mel features but yields a 0.4\% absolute improvement for the learnable front-end.
Applying SpecAugment in the \gls{STFT}-domain achieves the same absolute gain of 0.4\%.
In combination with larger time masks, the best result is obtained.
Interestingly, even the log Mel features benefit from masking in the \gls{STFT}-domain in combination with larger time masks.
Finally, feature masking based on the sorted order of learned filters does not improve the result.

\begin{table}[htbp]

\centering
\caption{Comparison of different SpecAugment variations. The given mask sizes for time (T) and frequency/feature (F) are upper limits for random sampling of the actual sizes. Results are reported on Hub5'00.}
\vspace{-1.5mm}
\label{table:results_specaug_variants_bs10k_v2}
\begin{tabular}{|c|c|c|c|c|}
\hline
\multirow{2}{*}{Features} & \multirow{2}{*}{SpecAugment} &           \multicolumn{2}{c|}{Mask Size} & \multirow{2}{*}{{WER [\%]}} \\\cline{3-4}
        \multirow{2}{*}{} &            \multirow{2}{*}{} &                   T &                  F &           \multirow{2}{*}{} \\\hline\hline
 \multirow{4}{*}{log Mel} &    \multirow{2}{*}{Baseline} &                  15 & \multirow{2}{*}{8} &                        12.8 \\\cline{3-3}\cline{5-5}
                          &                              &                  30 &                    &                        12.8 \\\cline{2-5}
                          & \multirow{2}{*}{STFT-domain} &                  15 &                  4 &                        12.8 \\\cline{3-5}
                          &                              &                  30 &                  8 &                        12.5 \\\hline
     \multirow{5}{*}{SCF} &    \multirow{2}{*}{Baseline} &                  15 &                 15 &                        13.2 \\\cline{3-5}
                          &                              &                  30 & \multirow{2}{*}{8} &                        12.8 \\\cline{2-3}\cline{5-5}
                          &                       Sorted & \multirow{2}{*}{15} &                    &                        13.3 \\\cline{2-2}\cline{4-5}
                          & \multirow{2}{*}{STFT-domain} &                     &                  4 &                        12.8 \\\cline{3-5}
                          &                              &                  30 &                  8 &                        12.6 \\
\hline
\end{tabular}

\end{table}

\begin{table}[htbp]

\centering
\vspace{-0.75mm}
\caption{Final results on Hub5'00 and Hub5'01 for the combination of the previously best performing audio perturbations (tempo perturbation) and SpecAugment settings.}
\vspace{-1.5mm}
\label{table:results_final_bs10k}
\begin{tabular}{|c|c|c|S[table-format=2.1]|S[table-format=2.1]|}
\hline
\multirow{2}{*}{Features} & \multirow{2}{*}{Perturbation} & \multirow{2}{*}{SpecAugment} & \multicolumn{2}{c|}{{WER [\%]}} \\\cline{4-5}
        \multirow{2}{*}{} &             \multirow{2}{*}{} &            \multirow{2}{*}{} & {Hub5'00} &           {Hub5'01} \\\hline\hline
 \multirow{3}{*}{log Mel} &                            no &     \multirow{2}{*}{default} &      12.8 &                11.5 \\\cline{2-2}\cline{4-5}
                          &          \multirow{2}{*}{yes} &                              &      12.4 &                11.2 \\\cline{3-5}
                          &                               &                  STFT-domain &      12.4 &                11.3 \\\hline
     \multirow{2}{*}{SCF} &                            no &                      default &      13.2 &                12.4 \\\cline{2-5}
                          &                           yes &                  STFT-domain &      12.5 &                11.3 \\
\hline
\end{tabular}

\end{table}

\subsection{Final Combination}
To obtain the final best model, we combine the best SpecAugment variation with the best audio perturbations.
We select tempo perturbation here as it performed best on both log Mel and learnable features.
While we also tried combining different perturbations, this resulted in an overall worse performance, likely because the regularization effect is too strong in this case.
\Cref{table:results_final_bs10k} presents the results.
As visible in the first three lines, tempo perturbation improves the log Mel baseline results on both Hub5'00 and Hub5'01.
In combination with tempo perturbation, SpecAugment in the \gls{STFT}-domain is not beneficial for log Mel features.
In contrast, the combination yields the best results for the learnable front-end which are now within 0.1\% absolute from the best log Mel results.
We also observe that the gap between train and dev scores is significantly smaller than in \Cref{fig:scores}, showing that overfitting has been mitigated effectively.
Therefore, we demonstrate that the presented regularization methods for training an \gls{ASR} system with learnable features allow closing the performance gap to traditional features.

\section{Conclusion}
This work focuses on regularization techniques for training \gls{CTC}-based \gls{ASR} models using both traditional and learnable feature extractors on the Switchboard dataset.
The examined audio perturbations can mitigate the overfitting during training and the overall improvements are relatively larger for learnable features.
Furthermore, we propose masking in the \gls{STFT}-domain as a simple but effective modification of SpecAugment to allow masking of contiguous frequency regions and ensuring variability in the input which was previously not the case for neural front-ends.
The final results with a combination of both techniques effectively close the performance gap between traditional and learnable features.

\ifinterspeechfinal
  \section{Acknowledgements}
  This work was partially supported by NeuroSys, which as part of the
initiative “Clusters4Future” is funded by the Federal Ministry of
Education and Research BMBF (funding ID 03ZU2106DD),
and by the project RESCALE within the program \textit{AI Lighthouse
Projects for the Environment, Climate, Nature and Resources} funded by
the Federal Ministry for the Environment, Nature Conservation, Nuclear
Safety and Consumer Protection (BMUV), funding ID: 6KI32006A.
\fi

\bibliographystyle{IEEEtran}
\bibliography{mybib}

\end{document}